\documentclass[sigconf]{acmart}
\AtBeginDocument{%
  }

\copyrightyear{2025}
\acmYear{2025}
\setcopyright{cc}
\setcctype{by}
\acmConference[MM '25]{Proceedings of the 33rd ACM International Conference on Multimedia}{October 27--31, 2025}{Dublin, Ireland}
\acmBooktitle{Proceedings of the 33rd ACM International Conference on Multimedia (MM '25), October 27--31, 2025, Dublin, Ireland}\acmDOI{10.1145/3746027.3758147}
\acmISBN{979-8-4007-2035-2/2025/10}

\usepackage{booktabs}
\usepackage{multirow}
\usepackage{enumitem}

\begin{document}

\title{Generative Flow Networks for Personalized Multimedia Systems: A Case Study on Short Video Feeds}

\author{Yili Jin}
\affiliation{%
  \institution{McGill University}
  \city{Montreal}
  \country{Canada}}

\author{Ling Pan}
\affiliation{%
  \institution{Hong Kong University of Science and Technology}
  \city{Hong Kong}
  \country{China}}

\author{Rui-Xiao Zhang}
\affiliation{%
  \institution{ByteDance Inc.}
  \city{San Diego}
  \country{USA}}

\author{Jiangchuan Liu}
\affiliation{%
  \institution{Simon Fraser University}
  \city{Burnaby}
  \country{Canada}}

\author{Xue Liu}
\authornote{Corresponding Author.}
\affiliation{%
  \institution{Mohamed bin Zayed University of Artificial Intelligence}
  \city{Abu Dhabi}
  \country{UAE}}
\affiliation{%
  \institution{McGill University}
  \city{Montreal}
  \country{Canada}}


\begin{abstract}
Multimedia systems underpin modern digital interactions, facilitating seamless integration and optimization of resources across diverse multimedia applications. To meet growing personalization demands, multimedia systems must efficiently manage competing resource needs, adaptive content, and user-specific data handling. This paper introduces Generative Flow Networks (GFlowNets, GFNs) as a brave new framework for enabling personalized multimedia systems. By integrating multi-candidate generative modeling with flow-based principles, GFlowNets offer a scalable and flexible solution for enhancing user-specific multimedia experiences. To illustrate the effectiveness of GFlowNets, we focus on short video feeds, a multimedia application characterized by high personalization demands and significant resource constraints, as a case study. Our proposed GFlowNet-based personalized feeds algorithm demonstrates superior performance compared to traditional rule-based and reinforcement learning methods across critical metrics, including video quality, resource utilization efficiency, and delivery cost.
Moreover, we propose a unified GFlowNet-based framework generalizable to other multimedia systems, highlighting its adaptability and wide-ranging applicability.
These findings underscore the potential of GFlowNets to advance personalized multimedia systems by addressing complex optimization challenges and supporting sophisticated multimedia application scenarios.
\end{abstract}





\maketitle
\section{Introduction}

\begin{figure*}
  \includegraphics[width=\textwidth]{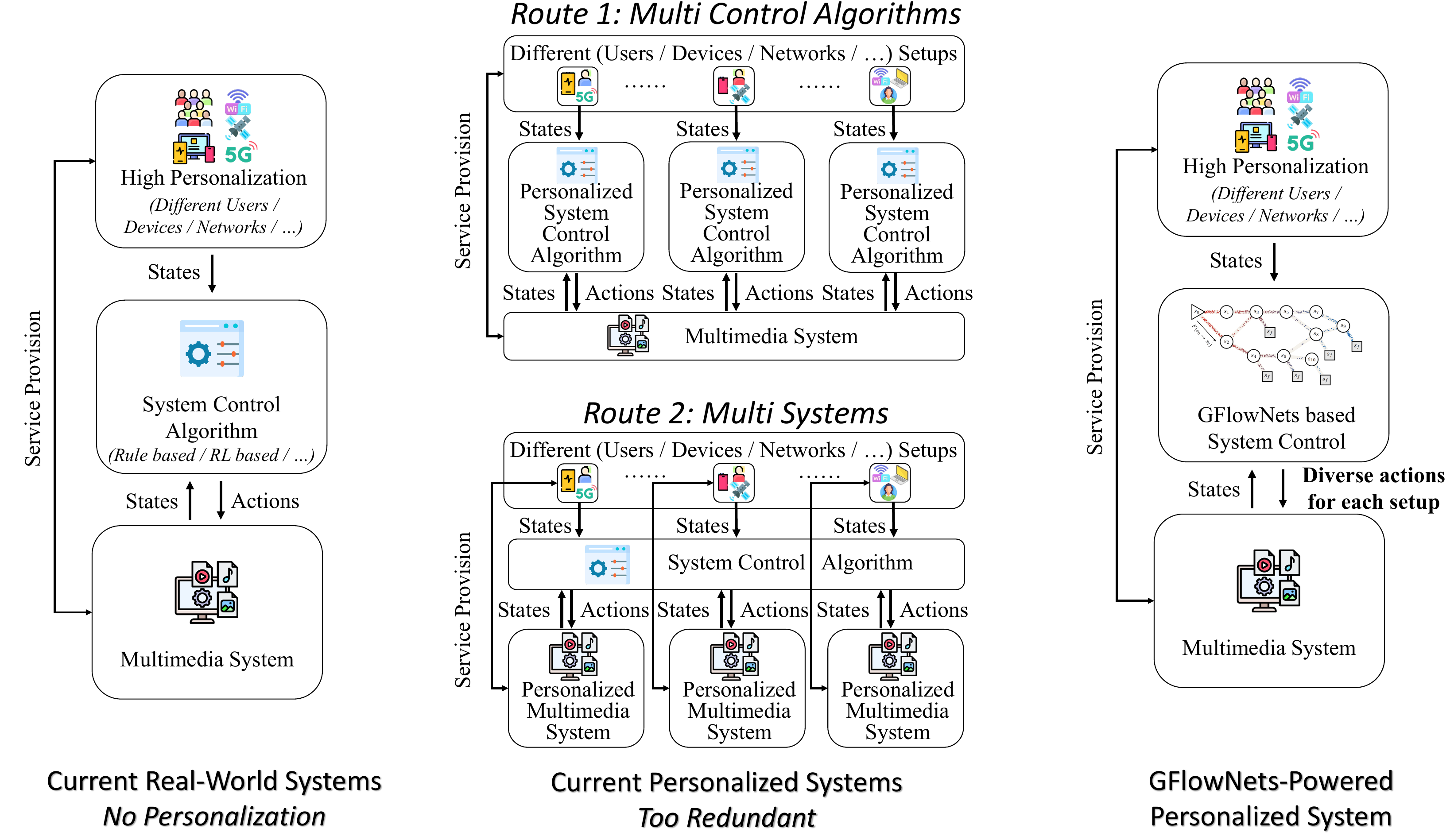}
  \vspace{-18pt}
  \caption{Frameworks of multimedia systems: (1) Current real-world systems without personalization; (2) Current personalized systems following two routes: multi control algorithms and multi systems; (3) GFlowNet-powered personalized system.}
  \vspace{-8pt}
  \label{fig:teaser}
\end{figure*}

In today’s digital era, multimedia systems form the backbone of engaging and highly interactive experiences, encompassing diverse applications including real-time multimedia communications, VR/AR, and multimedia-driven AI applications. These systems have become indispensable in various sectors including entertainment, education, healthcare, and enterprise, driving significant economic and social impacts. The ongoing evolution of multimedia systems is fueled by the growing demand for increasingly personalized content and user experiences~\cite{DBLP:conf/mm/ZhangJMDZF21,DBLP:conf/mm/PangZWHWLS18,DBLP:conf/mmsys/BhandarkarRD12,DBLP:conf/mmsys/YuDCLY17,DBLP:conf/mm/JinL0C22,DBLP:conf/mm/HuYJLCZ023}. Different users have varying preferences and access content through a wide array of devices, including smartphones, laptops, and VR/AR headsets. These devices, coupled with heterogeneous network environments such as Wi-Fi, 5G, and satellite networks, introduce unique performance constraints, latency requirements, and resource management challenges that multimedia systems must adapt to effectively.

Personalized short video feeds exemplifies these complexities and opportunities, emerging as a major driver of user engagement and network traffic globally. Short video feeds demands adaptive content delivery tailored precisely to individual user preferences, consumption patterns, and diverse network conditions. Managing these demands requires multimedia systems to dynamically balance critical factors such as video quality, buffering latency, bandwidth allocation, and computational resources, all while ensuring consistent user satisfaction.

However, current real-world multimedia systems often fall short in effectively addressing personalization. Most system control algorithms operate by simply reacting to system and user states to determine actions, without incorporating user-specific preferences. While research into personalization in multimedia systems has gained some attention, existing approaches generally follow two main technical routes: \textit{multi control algorithms}, and \textit{multi systems}. In the multi control algorithms route, control algorithms are fine-tuned for different personalized setups, for example, by adjusting parameters in rule-based algorithms or fine-tuning models in learning-based approaches to suit individual user setups. In contrast, the multi system route applies a common control algorithm but customizes the underlying system architecture for each user setup. An illustration of these frameworks are shown in Figure~\ref{fig:teaser}. While both routes offer personalization, they suffer from critical limitations. These include increased computational complexity, greater resource consumption, and scalability challenges, primarily due to the need for additional or specially tailored components that are explicitly designed to support personalization.

Generative Flow Networks (GFlowNets, GFNs), introduced by Bengio et al.~\cite{DBLP:conf/nips/BengioJKPB21,DBLP:journals/jmlr/BengioLDHTB23} in recent years, offer a brave new framework that combines multi-candidate generative modeling with flow-based programming principles. Initially created to facilitate efficient exploration and generation of diverse candidate solutions in active learning scenarios, GFlowNets exhibit significant potential to address personalization challenges in multimedia systems. Their unique capability to explore solution spaces adaptively and efficiently allows for a scalable, flexible approach to personalization, dynamically tailoring multimedia content and resource allocation precisely to individual system needs and user preferences.

In this paper, we present the first attempt to introduce GFlowNets as a foundation for enabling personalized multimedia systems. To demonstrate the capabilities and effectiveness of this novel framework, we present a detailed case study on personalized short video feeds. This choice is motivated by several factors: (1) video content constitutes the dominant form of multimedia traffic worldwide; (2) short video applications are immensely popular to daily life, and (3) short video feeds inherently exhibits highly personalized usage patterns. These characteristics make short video feeds a representative scenario to demonstrate the effectiveness of our framework. Our experimental results show that the GFlowNet-based approach significantly outperforms rule-based methods and reinforcement learning baselines across key metrics, including video quality, rebuffering rate, and bandwidth utilization, highlighting its potential for substantial impact in personalized multimedia applications.

Moreover, we propose and elaborate upon a unified GFlowNet-based framework that can be extended to broader multimedia system contexts. We detail how principles including adaptive state-action modeling, personalized reward function parameterization, and modular multi-candidate generation can be effectively adapted and applied to other critical multimedia domains such as multimedia transport protocols, multimedia data management and indexing techniques, and multimedia middleware infrastructure. This generalized approach not only emphasizes the versatility and broad applicability of GFlowNets but also facilitates their seamless integration into diverse multimedia systems, ensuring robust performance optimization and adaptability across various multimedia environments and applications.

The main contributions of this paper are as follows:

\begin{itemize}
\item From the multimedia application perspective, we propose a GFlowNet-based personalized short video feeds algorithm that achieves state-of-the-art performance.
\item From the multimedia systems perspective, we introduce a comprehensive GFlowNet-based framework for developing personalized multimedia applications, showcasing its flexibility and effectiveness across different environments.
\item From the GFlowNet perspective, we broaden its application domain by demonstrating its efficiency and effectiveness in personalized multimedia systems, extending its capabilities beyond previously explored use cases.
\end{itemize}

\section{Related Work}

In this section, we present an overview of the existing related works on GFlowNets and short video feeds.

\subsection{Generative Flow Network}

Generative Flow Networks (GFlowNets, GFNs), introduced by Bengio et al.~\cite{DBLP:conf/nips/BengioJKPB21,DBLP:journals/jmlr/BengioLDHTB23}, represent an advanced computational framework that merges generative modeling with flow-based programming principles. GFlowNets have primarily been explored in the context of generating diverse candidates for active learning applications. Their training objectives are designed to enable approximate sampling of outputs proportional to a specified reward function, thereby improving the efficiency and effectiveness of the learning process. Some recent works~\cite{DBLP:conf/icml/KimKYZPKPBB24,DBLP:conf/iclr/PanJMB24,DBLP:conf/mm/ZhuL0HWK0W23} further improve its performance.

To date, the application of GFlowNets has been predominantly concentrated in the field of biology~\cite{DBLP:conf/icml/JainBHRDEFZKZSD22,DBLP:conf/nips/ZhuWHYhHW23}. However, their potential reaches far beyond these domains. We posit that the adaptive and dynamic nature of GFlowNets makes them particularly well-suited for multimedia systems, especially in personalized services.

To the best of our knowledge, this paper is the first to explore the application of GFlowNets in multimedia area, specifically addressing the challenges of personalized systems. As a case study, we demonstrate their prioritization capabilities in short video feeds, showcasing the versatility and scalability of GFlowNets. This novel application opens new avenues for research and highlights their promise in addressing emerging challenges in multimedia area.

\subsection{Short Video Feeds}
The rise in popularity of user-generated, short-form video sharing on social media platforms has led to significant bandwidth costs. Key challenges include reducing bandwidth usage without impacting the quality of user experience~\cite{DBLP:conf/mm/HuangZJZS22,DBLP:conf/mm/ZuoLXOLJZ0022,DBLP:conf/mm/WuZC22}. In these platforms, users frequently swipe through videos, often skipping to the next before the current one finishes. This behavior necessitates adaptive and strategic prefetching of videos to prevent bandwidth wastage.

Adaptive prefetching faces complications due to unpredictable user behaviors, influenced by individual content preferences and viewing histories, and the competition among multiple videos for bandwidth. Decisions on which videos to prefetch and at what bitrate, using adaptive bitrate algorithms~\cite{DBLP:conf/sigcomm/HuangJMTW14,DBLP:journals/comsur/SaniME17}, need to adapt to variable network conditions.

Prior research on video-on-demand focused on dynamic prefetching for single videos~\cite{DBLP:journals/tmm/ChenZC15}, which is less applicable in multiple short video scenarios. Recent advancements~\cite{DBLP:conf/nossdav/HeHZW20,DBLP:conf/icmcs/Li0MYX22} have explored more dynamic strategies that aim to reduce rebuffering and bandwidth wastage. A wastage-aware approach~\cite{DBLP:journals/tmc/ZhangLGL23} is proposed to minimize data wastage based on predictable network conditions and user behaviors. The use of diverse user behavior data for adaptive video prefetching shows potential for future improvements.

Despite these advancements, personalized approaches to short video feeds, which adapt to individual user preferences and network environments, remain underexplored. Addressing this gap could unlock new opportunities for enhancing user experience and resource efficiency.

\section{Short Video Feeds Formulation} \label{sec:formulation}

In this section, we formulate the problem of short video feeds services, beginning with a typical framework, defining the objective and decision space, and highlighting the personalization challenges.

\subsection{Framework}

\begin{figure}[t]
    \centering
    \includegraphics[width=0.7\linewidth]{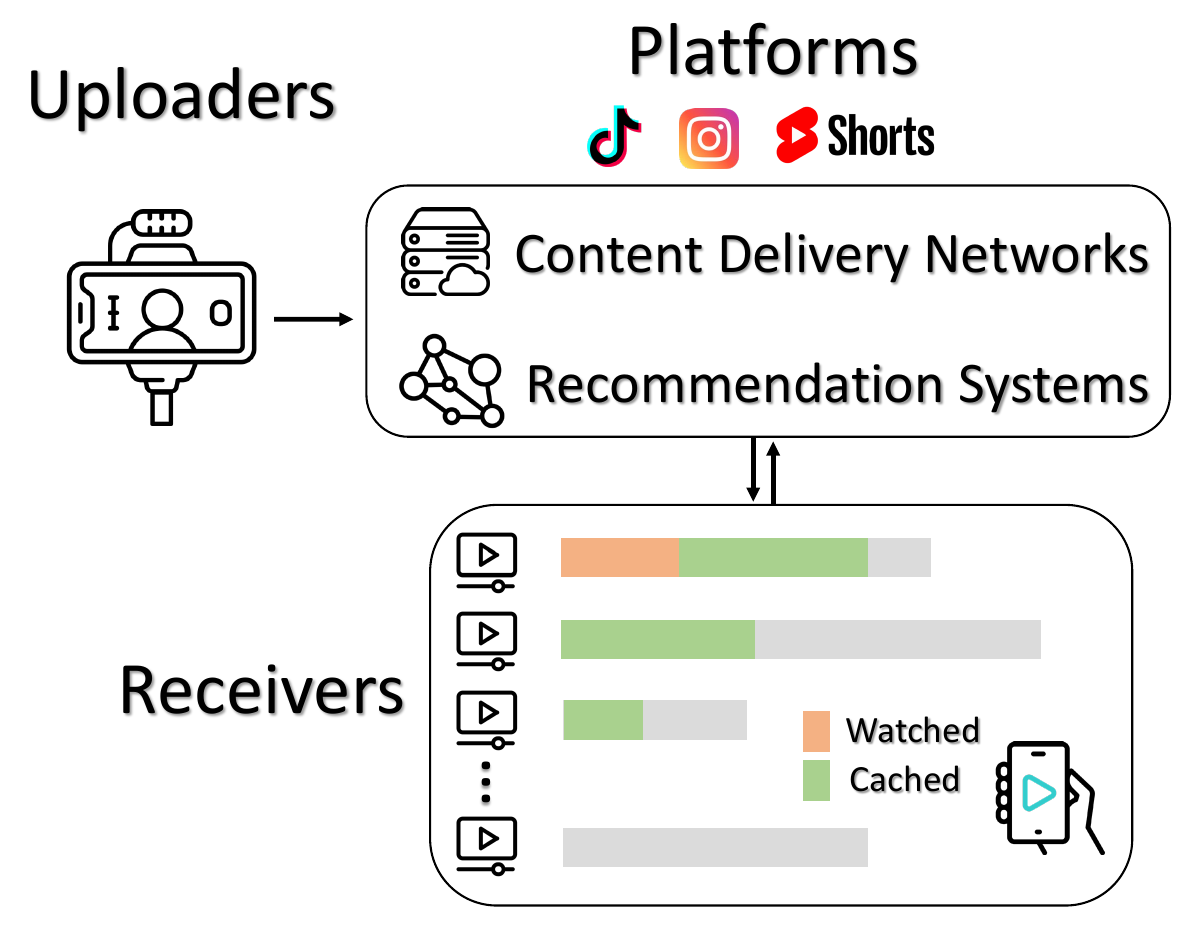}
    \caption{The framework of short video feeds service.}
    \label{fig:framework}
\end{figure}

Figure~\ref{fig:framework} illustrates the framework of a typical short video feeds service. Uploaders create and upload short videos to the platform, where two key components handle the feeds process: Content Delivery Networks and Recommendation Systems. Content Delivery Networks are responsible for efficiently delivering and streaming videos from storage to users, while Recommendation Systems determine which videos should be delivered.
On the receiver side, the client device caches certain video chunks in advance. Unlike traditional video streaming services, short video service allows users to swipe to the next video at any time. This unique behavior requires the system to cache not only the current video but also videos from the recommendation list. Furthermore, the system must dynamically adjust the bitrate of cached videos based on the current network conditions. The goal is to optimize the user's Quality of Experience (QoE) while minimizing bandwidth costs.

\subsection{Objective}

Firstly, we define the user's QoE using a widely-adopted QoE model in the video streaming domain~\cite{DBLP:conf/sigcomm/YinJSS15,DBLP:conf/sigcomm/MaoNA17}\footnote{It may also be referred to as QoS (Quality of Service), since some definitions of QoE (Quality of Experience) include only perception metrics~\cite{Qualinet}. Here, we adopt the definition from the video streaming domain and refer to it as QoE.}:
\begin{equation}
\text{QoE} = \alpha \sum_{n=1}^{N} q(B_n) - \beta \sum_{n=1}^{N} T_n - \gamma \sum_{n=1}^{N-1} \left| q(B_{n+1}) - q(B_n) \right|,
\end{equation}
where $N$ represents the total number of video chunks watched. $B_n$ is the bitrate of chunk $n$, and $q(B_n)$ maps that bitrate to the quality perceived by a user. $T_n$ denotes the rebuffering time caused by downloading chunk $n$ at bitrate $B_n$. The final term penalizes fluctuations in video quality to promote smooth playback. The coefficients $\alpha$, $\beta$, and $\gamma$ are weights representing the relative importance of bitrate, rebuffering time, and smoothness, respectively. Different users may have varying preferences, leading to different weights.

Next, we define the bandwidth cost (BW) as:
\begin{equation}
\text{BW} = \sum_{m=1}^{M} S_m,
\end{equation}
where $M$ represents the total number of video chunks downloaded. $S_m$ is the size of chunk $m$.

The objective is then defined as:
\begin{equation}
\text{maximize} \ \text{QoE} - \theta \ \text{BW},
\end{equation}
where $\theta$ is a weight that balances QoE and bandwidth cost. Similar to QoE weights, different users may have unique preferences, resulting in different values for $\theta$.

\subsection{Design Space}

Optimizing user QoE in short video feeds requires careful prefetching of both the currently viewed video and the videos in the recommendation queue. However, when a user quickly swipes past a video, any prefetched but unwatched data results in wasted bandwidth. To address this, the algorithm must dynamically decide which video chunks to prefetch and at what bitrate, based on real-time network conditions and the buffer states of the videos.

Moreover, to minimize bandwidth waste, the algorithm may pause prefetching during favorable network conditions to avoid unnecessary data downloads. Determining the optimal duration for such pauses is another critical consideration for reducing inefficiencies while maintaining a seamless viewing experience.

At each decision point, the algorithm must answer three key questions: whether to prefetch a video, which video to prefetch, and at what bitrate to prefetch it.

\subsection{Personalization Challenge}

Personalization in short video feeds is challenging due to diverse user preferences, dynamic behaviors, and contextual variability. Users prioritize different aspects of QoE, such as video quality, smoothness, or rebuffering, requiring tailored adjustments of the weights $\alpha$, $\beta$, and $\gamma$. Behavior is unpredictable, with users frequently swiping or rewatching videos, complicating content prefetching. Contextual factors, such as network conditions or device type, further influence preferences, making static optimization ineffective.
Additionally, balancing QoE and bandwidth cost is user-specific, as the trade-off varies based on network access and willingness to consume data. The parameter $\theta$ must reflect these individual differences. 

\section{GFlowNet-Based Approach}

In this section, we present our GFlowNet-based approach for personalized short video feeds, detailing how GFlowNets can be adapted to address diverse user preferences and demonstrating the end-to-end pipeline.

\subsection{GFlowNets Design}

Consider a directed acyclic graph (DAG) $\mathcal{G} = (\mathcal{S}, \mathcal{A})$, where $\mathcal{S}$ denotes the set of states and $\mathcal{A}$ the set of actions. Our objective is to learn a stochastic policy $\pi$ that constructs discrete objects $x \in \mathcal{X}$ with probability proportional to a reward function $R(x)$, that is, $\pi(x) \propto R(x)$. The agent proceeds by selecting actions to move from one state to another until it reaches a terminal state. This sequence of states is described by a trajectory $\tau = (s_0, \dots, s_n) \in \mathcal{T}$, where $\mathcal{T}$ is the collection of all possible trajectories.

Following \cite{DBLP:conf/nips/BengioJKPB21,DBLP:journals/jmlr/BengioLDHTB23}, we define a trajectory flow $F(\tau)$, which is a non-negative quantity assigned to each trajectory. The state flow $F(s)$, representing the sum of flows for every trajectory passing through a state $s$, is given by
\begin{equation}
F(s) = \sum_{\tau \ni s} F(\tau).
\end{equation}
Similarly, the edge flow $F(s \to s')$ is defined as the total flow for all trajectories that include the transition from $s$ to $s'$, namely
\begin{equation}
F(s \to s') = \sum_{\tau \ni s \to s'} F(\tau).
\end{equation}
Then, we introduce the forward policy
\begin{equation}
P_F(s' \mid s) = \frac{F(s \to s')}{F(s)},
\end{equation}
which indicates the probability of moving from $s$ to $s'$, and the backward policy
\begin{equation}
P_B(s \mid s') = \frac{F(s \to s')}{F(s')},
\end{equation}
which captures the probability of returning from $s'$ to $s$.

A flow is called consistent if, for each state, the total incoming flow equals the total outgoing flow. When flow consistency holds, sampling from this policy produces objects $x$ in proportion to $R(x)$, thereby matching the reward-based distribution.

Flow Matching (FM) provides a learning objective by parameterizing the edge flow function \(F_{\theta}(s, s')\) with learnable parameters \(\theta\). The FM loss is defined for each state \(s\) as:
\begin{equation}
\mathcal{L}_{\text{FM}}(s) = \biggl( \log \sum_{s'' \to s} F_{\theta}(s'', s) -
\log \sum_{s \to s'} F_{\theta}(s, s') \biggr)^2.
\end{equation}
For a terminal state \(s\), the outgoing flow term is replaced by its reward \(R(s)\). Optimization is performed in log-space for numerical stability.

Trajectory Balance (TB) introduces an alternative objective based on detailed balance conditions, minimizing the following loss for each trajectory $\tau$:
\begin{equation}
\begin{aligned}
\mathcal{L}_{\text{TB}}(\tau) 
&= \Bigl(\log Z_{\theta}\,\prod_{t=0}^{n-1} P_{F_{\theta}}(s_{t+1} \mid s_t) \\
&\quad-\;\log R(x)\,\prod_{t=0}^{n-1} P_{B_{\theta}}(s_t \mid s_{t+1})\Bigr)^2,
\end{aligned}
\end{equation}
where \(Z_{\theta}\) is a learnable partition function, and \(x\) is the terminal object represented by \(s_n\).

Finally, the Detailed Balance (DB) condition enforces flow consistency at every state transition:
\begin{equation}
F(s) P_F(s' \mid s) = F(s') P_B(s \mid s'), \ \forall (s\to s') \in \mathcal{A}.
\end{equation}
For terminal states \(x\), the flow must satisfy \(F(x) = R(x)\). When DB is satisfied for all transitions, the learned policy accurately samples objects in proportion to \(R(x)\).

\subsection{Pipeline}

\begin{figure}[t]
    \centering
    \includegraphics[width=1\linewidth]{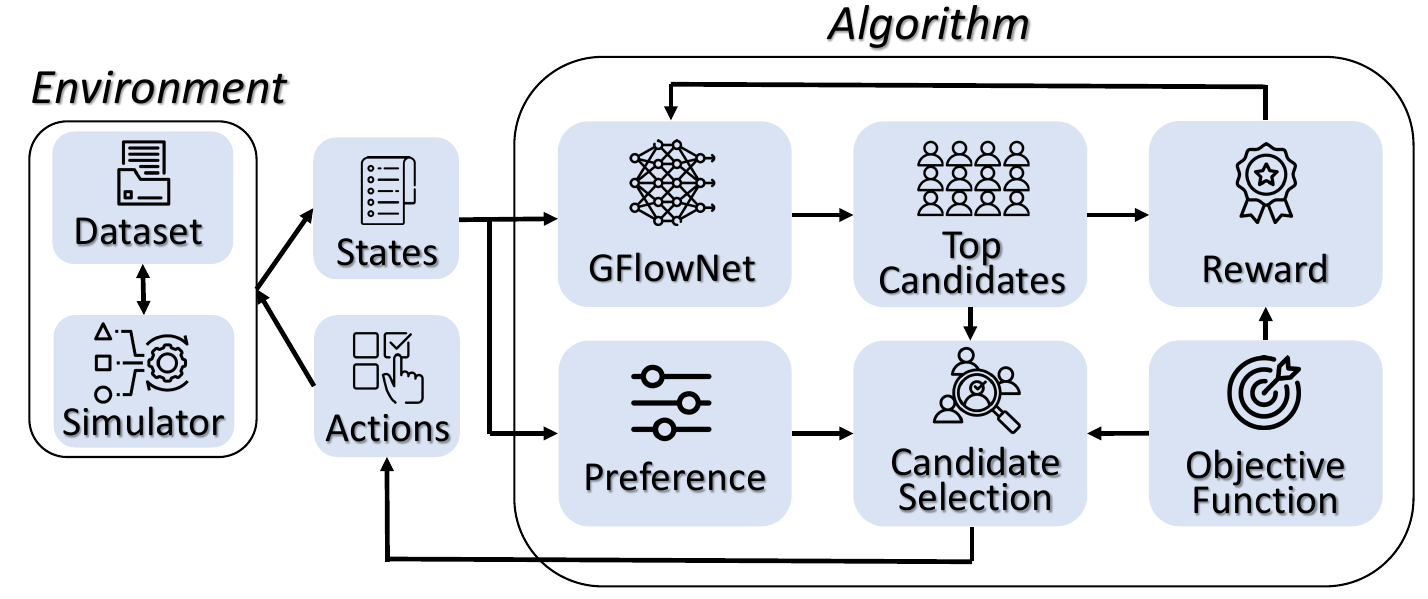}
    \caption{Pipeline of our proposed GFlowNet-based approach.}
    \label{fig:algorithm}
\end{figure}

Figure~\ref{fig:algorithm} illustrates the overall pipeline of our proposed GFlowNet-based algorithm for personalized short video feeds. The environment, represented here by a combination of a pre-collected dataset and a simulator in our experiments, could also be substituted with a real-world environment. The environment outputs a \emph{state} consisting of video-related information (e.g., video chunk size, the bitrate of cached segments), network-related information (e.g., rebuffering time, bandwidth), and user preference parameters. The algorithm then determines an \emph{action}, which specifies which video chunk to download next and at which bitrate. Section~\ref{sec:formulation} outlines the formulation of these elements.

Our algorithm follows an RL-like learning paradigm but leverages the expressive capacity of GFlowNets:

\begin{enumerate}
    \item We first feed the current state into a GFlowNet, which generates a ranked list (e.g., the top 10) of candidate actions.
    \item For each candidate action, we compute the \emph{objective value} based on user preference.
    \item We then select the single best candidate action (i.e., the one maximizing the objective) to execute in the environment.
    \item To train the GFlowNet, we average the objective values of all 10 candidates to form a reward signal. This reward is used to update the GFlowNet parameters, encouraging it to generate high-quality candidate actions in future steps.
\end{enumerate}

The key advantage lies in the GFlowNet’s ability to propose multiple promising candidate actions at every decision point. Traditional approaches typically produce a single action or a single distribution over actions. In contrast, by generating a \emph{set} of candidates, we can explicitly incorporate user preferences and diverse environmental conditions to pick the most appropriate action for each user at each step. 

Furthermore, because the training reward is based on the \emph{average} performance of the candidates rather than just the single chosen action, the GFlowNet is encouraged to maintain a broad \emph{distribution} of good solutions. This design makes it naturally flexible for handling different user preferences: as these preferences change (e.g., placing more emphasis on bitrate quality or rebuffering time), the GFlowNet’s sampling distribution can smoothly adapt, ensuring that personalized requirements are consistently met.

Hence, this pipeline offers both \emph{personalization} and \emph{robustness} by integrating the strengths of generative modeling (through GFlowNets) with traditional RL objectives (through reward maximization). In multimedia system scenarios, where user behavior and preferences can be highly variable, this approach provides a powerful mechanism for dynamically tailoring QoE metrics and bandwidth utilization to each user’s individual demands.

\section{Experiments}

In this section, we present a comprehensive evaluation of our proposed GFlowNet-based short video feeds algorithm.

\subsection{Setups}

\subsubsection{Simulator}

We use a chunk-level simulator, adapted from \cite{DBLP:conf/mm/ZuoLXOLJZ0022}, to emulate short video feeds and evaluate our algorithms. The simulator focuses on two main processes: downloading video chunks and playing the currently viewed video. It takes as input four types of data, network traces, video traces, user traces, and user preference parameters. Those data cover bandwidth conditions, chunk-level sizes at multiple bitrates, user watch-time behaviors, and individual user preferences.
Unlike traditional single-buffer approaches, this simulator maintains a separate buffer for each video in the recommendation queue. For each downloaded chunk, it calculates the download time based on the current bandwidth and chunk size, updates the corresponding buffer, and tracks rebuffering events that occur when a user attempts to play a video with insufficient buffered data. Rebuffering is often triggered by users swiping rapidly through multiple videos, exhausting buffers faster than downloads can complete. The simulator also supports a pause mechanism in which downloads halt but video playback continues. This feature allows algorithms to strategically decide when to pause in order to balance QoE and bandwidth usage.

\subsubsection{Datasets}

First, the network traces come from \cite{DBLP:conf/nossdav/LiZZCM23}, which consolidates publicly available logs from Oboe~\cite{DBLP:conf/sigcomm/AkhtarNGRCKRZZ18} and FCC~\cite{fcc}, and categorizes them into low, medium, or high bandwidth conditions using thresholds of 1.5 and 3~Mbps. Second, the video and user datasets are drawn from DUASVS~\cite{DBLP:journals/tsc/ZhangZLGLHA23}, containing millions of records documenting per-chunk statistics and user retention rates. Finally, we synthesize user preference parameters (i.e., $\alpha$, $\beta$, $\gamma$, and $\theta$) for each user. Each simulation scenario is formed by selecting one network trace, a sequence of short videos, and a user’s retention profile. This combined setup enables robust and realistic evaluations of how well different algorithms adapt to variations in network conditions, video characteristics, and user preferences.

\subsubsection{Baselines}

\begin{table*}[t]
\centering
\caption{Overall performance under different bandwidth conditions. All QoE values are normalized to [0,1] (higher is better). Rebuffering is measured in seconds (lower is better), and Bandwidth Wastage is expressed as a percentage.}
\vspace{-5pt}
\begin{tabular}{@{}lccccc@{}}
\toprule
\textbf{Method} & \textbf{Scenario} & \textbf{QoE} & \textbf{Rebuf (s)} & \textbf{BW Usage (MB)} & \textbf{BW Wastage (\%)} \\
\midrule
\multirow{3}{*}{PDAS}     
   & Low    & 0.65  &  5.71  &  62.2  & 20.8 \\
   & Medium & 0.72  &  4.01  &  68.4  & 17.6 \\
   & High   & 0.76  &  3.47  &  79.1  & 15.9 \\
\midrule
\multirow{3}{*}{DAM}
   & Low    & 0.67  &  5.11  &  58.7  & 19.3 \\
   & Medium & 0.75  &  3.89  &  66.8  & 16.8 \\
   & High   & 0.79  &  2.95  &  77.2  & 15.2 \\
\midrule
\multirow{3}{*}{Incendio}
   & Low    & 0.70  &  4.64  &  56.9  & 18.4 \\
   & Medium & 0.78  &  3.52  &  65.2  & 15.1 \\
   & High   & 0.83  &  2.43  &  74.5  & 13.5 \\
\midrule
\multirow{3}{*}{\textbf{Ours}} 
   & Low    & 0.80  &  4.05  &  55.1  & 16.2 \\
   & Medium & 0.88  &  2.97  &  63.9  & 13.9 \\
   & High   & 0.94  &  2.05  &  72.3  & 12.7 \\
\bottomrule
\end{tabular}
\label{tab:overall}
\end{table*}

We choose three representative baselines:

\begin{itemize}
\item \textbf{DAM}~\cite{DBLP:conf/mm/QianXPZL22}. DAM is a basic deep reinforcement learning approach that jointly determines the video to play, the bitrate level, and any sleep/pause duration.
\item \textbf{PDAS}~\cite{DBLP:conf/mm/ZhouBZGY22}. Proposed by Kuaishou, PDAS is a state-of-the-art rule-based approach that maximizes the same utility score by combining a user retention model with a hand-designed buffer management strategy. It also leverages RobustMPC~\cite{DBLP:conf/sigcomm/YinJSS15} to optimize decisions over a multi-chunk horizon, using buffer occupancy and throughput predictions.
\item \textbf{Incendio}~\cite{DBLP:conf/nossdav/LiZZCM23}. It is a state-of-the-art learning-based approach that applies Multi-Agent Reinforcement Learning with expert guidance, separating decisions for the video choice and video bitrate into respective buffer management and bitrate adaptation agents.
\end{itemize}

\subsection{Overall Performance}

Table~\ref{tab:overall} summarizes the performance of our proposed GFlowNet-based personalized feeds algorithm and the three baselines under three different network conditions (Low, Medium, and High). We report the following metrics:
\begin{itemize}
    \item \textit{Normalized QoE}: A QoE score normalized to the range [0,1], which integrates the perceived video quality with rebuffering and smoothness penalties.
    \item \textit{Rebuffering Time}: The total rebuffering duration (in seconds) per video session.
    \item \textit{Bandwidth Usage}: The total data downloaded (in MB).
    \item \textit{Bandwidth Wastage}: The percentage of data downloaded but never viewed (e.g., due to user swiping).
\end{itemize}

As demonstrated, our approach consistently outperforms state-of-the-art baselines. In low bandwidth, it enhances QoE by approximately 14\% while reducing bandwidth usage by 3\%. In medium bandwidth, it improves QoE by around 13\% and lowers bandwidth consumption by 2\%. Under high bandwidth, our method achieves a 13\% QoE improvement while reducing bandwidth usage by approximately 3\%.

These results highlight the effectiveness of our method in delivering a superior user experience with higher QoE while optimizing resource utilization by reducing bandwidth consumption across diverse network conditions. Additionally, our method minimizes rebuffering time and reduces bandwidth wastage. These improvements suggest that by leveraging a multi-candidate policy that explicitly accounts for user retention probabilities, GFlowNet can make more adaptive and efficient prefetching decisions tailored to varying network conditions and user behaviors.

\begin{table}[t]
\centering
\caption{Ablation study on candidate generation.}
\vspace{-5pt}
\begin{tabular}{@{}lccc@{}}
\toprule
\textbf{Variant} & \textbf{QoE} & \textbf{Rebuf (s)} & \textbf{BW Wastage (\%)} \\
\midrule
SC & 0.76  & 3.42  & 15.8 \\
MC & \textbf{0.88}  & \textbf{2.97}  & \textbf{13.9} \\
\bottomrule
\end{tabular}
\label{tab:ablation}
\end{table}

\subsection{Impact of Multi-Candidate Generation}

To further investigate the contributions of our design, we analyze the effect of generating multiple candidate actions per decision step, a key feature that enables GFlowNet to support personalization. We compare two variants of our approach:
\begin{itemize}
    \item \textbf{Single-Candidate (SC):} The model outputs only one candidate action per decision, similar to conventional reinforcement learning policies.
    \item \textbf{Multi-Candidate (MC):} The full GFlowNet implementation generates a ranked list of candidate actions, from which the best is selected.
\end{itemize}

Table~\ref{tab:ablation} presents the ablation results, demonstrating the effectiveness of the MC mechanism. Compared to the SC approach, MC not only improves QoE from 0.76 to 0.88 but also reduces rebuffering time from 3.42 seconds to 2.97 seconds. Additionally, MC decreases bandwidth wastage from 15.8\% to 13.9\%, highlighting its efficiency in optimizing resource utilization while enhancing user experience.

This ablation study clearly shows that the multi-candidate generation mechanism is critical for achieving better performance. By exploring a broader set of candidate actions at every decision point, our approach can better adapt to varying network conditions and user behaviors, resulting in reduced rebuffering and lower bandwidth wastage.

\subsection{Impact of Personalization}

\begin{table}[t]
\centering
\caption{Impact of Personalization: Comparison between the non-personalized and personalized variants.}
\vspace{-5pt}
\begin{tabular}{@{}lccc@{}}
\toprule
\textbf{Variant} & \textbf{QoE} & \textbf{Rebuf (s)} & \textbf{BW Wastage (\%)} \\
\midrule
Fixed Preferences & 0.78 & 3.52 & 14.4 \\
Personalized       & \textbf{0.88} & \textbf{2.97} & \textbf{13.9} \\
\bottomrule
\end{tabular}
\label{tab:personalization}
\end{table}

To further quantify the benefits of personalization, we compare our full personalized approach against a non-personalized variant. In the non-personalized variant, all users are assigned a fixed set of preference parameters, rather than tailoring them individually.

Table~\ref{tab:personalization} presents the impact of personalization, demonstrating its significant benefits. Compared to the fixed-preference approach, the personalized variant improves QoE from 0.78 to 0.88, reduces rebuffering time from 3.52 seconds to 2.97 seconds, and lowers bandwidth wastage from 14.4\% to 13.9\%.

These results highlight that by tailoring prefetching decisions to individual user preferences, our GFlowNet approach more effectively balances video quality, playback smoothness, and bandwidth efficiency. In dynamic user scenarios, such personalization is crucial for optimizing the overall feeds experience.

\section{A Unified GFlowNet-based Framework for Other Multimedia Systems}

While this paper uses personalized short video feeds as a primary case study, our proposed GFlowNet-based framework can be naturally extended to a broad spectrum of multimedia systems. The fundamental approach, leveraging generative modeling techniques to explore diverse candidate solutions dynamically and efficiently, accounting for user-specific preferences and contextual parameters, can be applied effectively across various multimedia domains. 

For example, in scalable multimedia data management systems~\cite{DBLP:journals/cacm/Grosky97,DBLP:conf/tpsisa/Thuraisingham23}, efficient browsing, retrieval, and recommendation mechanisms are critical for handling extensive multimedia content databases. GFlowNets can generate adaptive indexing structures and personalized query optimization strategies, efficiently balancing retrieval accuracy, computational overhead, and user satisfaction.
Moreover, multimedia middleware platforms~\cite{DBLP:conf/mm/DukeH98,DBLP:journals/expert/ZhouXSVY10,DBLP:conf/tvx/BoaroCMSRDSC24} that manage complex interactions between diverse system components such as storage, computational resources, and user interface modules can greatly benefit from our approach. By dynamically generating and evaluating multiple candidate resource allocation and scheduling policies, GFlowNets provide robust mechanisms to ensure optimal resource utilization, reduced latency, and enhanced user experience.
Other promising application scenarios include but not limit to adaptive multimedia content encoding and transcoding strategies~\cite{DBLP:journals/tcsv/VetroSW01,DBLP:journals/twc/WangBZ22,DBLP:journals/tomccap/JinLLL25} that cater to diverse device capabilities and network conditions, and real-time multimedia analytics systems~\cite{DBLP:journals/tomccap/ZhangYZE19a,DBLP:journals/tist/NieZWHFC17} that require adaptive data processing pipelines to balance computational resource constraints with analytics accuracy. 

The inherent adaptability and scalability of the GFlowNet framework provide a versatile tool capable of significantly enhancing performance and personalization across a wide variety of multimedia systems.
Below, we present a unified framework for integrating our proposed GFlowNet-based approach into various personalized multimedia systems.

\vspace{-5pt}\paragraph{1. Unified State-Action Modeling.}  
At the heart of any multimedia system application is the need to capture system dynamics and user context within a well-defined state space. Let the state \( s \in \mathcal{S} \) be a comprehensive feature vector that includes both system metrics (e.g., throughput, latency, buffer occupancy, resource utilization) and user-specific information (e.g., preference profiles, historical behavior patterns). Similarly, the action \( a \in \mathcal{A} \) is defined to encompass all feasible control variables or policy adjustments (such as routing decisions, scheduling configurations, or resource allocation levels). The state-action pair is formally represented as:
\begin{equation}
    (s, a) \in \mathcal{S} \times \mathcal{A}.
\end{equation}
This abstraction ensures that the framework is agnostic to the specific multimedia application and can be tailored to a wide array of domains.

\vspace{-5pt}\paragraph{2. Reward Function Parameterization.}  
A central component of our framework is the reward function \( R(s, a) \), which is engineered to encapsulate multiple performance objectives, such as Quality-of-Service (QoS), Quality-of-Experience (QoE), energy efficiency, and cost. To incorporate personalization, the reward function is parameterized with user-specific weights. A general form of the reward function may be expressed as:
\begin{equation}
    R(s, a) = \alpha\, Q(s, a) - \beta\, L(s, a) - \gamma\, C(s, a),
\end{equation}
where:
\begin{itemize}
    \item \(Q(s, a)\) represents quality metrics (e.g., multimedia data quality, throughput),
    \item \(L(s, a)\) denotes latency or delay, and
    \item \(C(s, a)\) captures the cost or energy consumption associated with the action \(a\).
\end{itemize}
The weights \(\alpha\), \(\beta\), and \(\gamma\) can be tuned to reflect individual user preferences and overall system requirements.

\vspace{-5pt}\paragraph{3. Trajectory-Based Policy Learning via GFlowNets.}  
Our approach leverages the intrinsic capability of GFlowNets to learn a stochastic policy that generates complete trajectories through the state-action space. A trajectory is defined as
\begin{equation}
    \tau = (s_0, a_0, s_1, a_1, \dots, s_n),
\end{equation}
with an associated cumulative reward:
\begin{equation}
    R(\tau) = \sum_{t=0}^{n-1} R(s_t, a_t).
\end{equation}
GFlowNets aim to sample trajectories with probability proportional to their rewards, i.e.,
\begin{equation}
    P(\tau) \propto R(\tau).
\end{equation}
Training objectives such as Flow Matching and Trajectory Balance are employed to ensure this proportionality, thereby enabling the exploration of a rich and diverse set of candidate policies for complex, multi-objective multimedia tasks.

\vspace{-5pt}\paragraph{4. Modular Multi-Candidate Generation and Selection.}  
A key advantage of our framework is its ability to generate multiple candidate actions at each decision step. In practice, the GFlowNet produces a ranked list of candidate trajectories \( \{\tau_1, \tau_2, \dots, \tau_k\} \), each corresponding to a distinct system control strategy. A selection module then evaluates these candidates, using either a learned estimator or domain-specific heuristics, to choose the one that maximizes the expected reward:
\begin{equation}
    \tau^* = \arg\max_{\tau_i} \; \mathbb{E}[R(\tau_i)].
\end{equation}
This separation between candidate generation and candidate selection not only increases robustness but also simplifies integration with existing system controllers.

\vspace{-5pt}\paragraph{5. Integration and End-to-End Adaptation.}  
The proposed framework is designed to function as an intelligent middleware layer between raw system measurements and traditional control mechanisms. By continuously updating its state representation with real-time data and re-evaluating the candidate actions, the framework maintains an adaptive policy that responds to dynamic system conditions and evolving user behavior. The end-to-end learning paradigm inherent in GFlowNets ensures that the entire process, from state sensing to action execution, is optimized jointly, allowing the framework to autonomously improve over time.

In summary, by systematically designing the state-action representation, reward function, and candidate generation modules, our GFlowNet-based framework provides a scalable pathway for integrating advanced AI methods into next-generation multimedia systems. This unified, technical approach ultimately enables adaptive, user-centric performance optimization across diverse multimedia applications.

\section{Discussion}

This paper presents a new perspective on personalized multimedia systems through the application of GFlowNets. In this section, we critically reflect on the limitations of our work and propose directions for future research.

\subsection{Limitations}

While our GFlowNet-based approach demonstrates promising performance in simulated short video feeds, several limitations warrant further investigation. First, our evaluation relies on synthesized user preferences and controlled simulation environments, which may not fully capture the variability and complexity of real-world network conditions and user behaviors. Second, although our multi-candidate generation mechanism does not introduce additional latency, its effectiveness depends on the robustness of the candidate selection module and the precision of the underlying user behavior and network models. Finally, the performance of our approach is closely tied to the design of the reward function; any mis-specification of reward parameters may limit its personalization benefits across diverse settings.

\subsection{Future Research Directions}

The conceptual shift proposed in this paper, from deterministic personalization policies to generative candidate-based reasoning, opens several promising research directions:

\vspace{-5pt}\paragraph{Personalization via Generative Distributions.}
Traditional approaches to personalization aim to learn the “best” decision for each user. Our work shifts the focus toward sampling from a high-quality distribution of possible decisions, implicitly encoding the diversity of user preferences and system contexts. This opens up novel opportunities to study personalization under uncertainty, multi-user fairness, and controllable diversity, especially relevant for shared or collaborative multimedia systems.

\vspace{-5pt}\paragraph{GFlowNets for Long-Horizon Planning}
Our current GFlowNets focuses on one-step decision-making. However, their generative nature naturally supports sequential decision-making under long-term objectives. This could be highly beneficial for multimedia tasks requiring temporally extended reasoning, such as progressive content rendering, end-to-end delivery scheduling, or story-aware video recommendation systems. Incorporating temporal credit assignment and structured planning modules into the GFlowNet pipeline is a promising area for future work.

\vspace{-5pt}\paragraph{Reward Learning and Human-in-the-Loop Systems}
Given the challenge of specifying reward functions that accurately reflect human intent, integrating users more explicitly into the loop is a natural next step. GFlowNets could be coupled with preference elicitation methods, such as pairwise comparisons or natural language feedback, to iteratively refine their reward models. This approach aligns with recent trends in AI safety~\cite{DBLP:journals/corr/abs-2501-17805}, and could lead to more transparent, controllable, and trustworthy multimedia systems.

\vspace{-5pt}\paragraph{Cross-Domain Applicability}
Although we focus on short video feeds, the framework proposed is domain-agnostic. We foresee potential applications in many multimedia domains. Each of these domains presents its own optimization trade-offs and system constraints, making them fertile ground for generalizing and stress-testing the GFlowNet-based paradigm.

\section{Conclusion}

This paper presents a brave new framework for personalized multimedia systems grounded in GFlowNets, with short video feeds as a representative use case. By leveraging the generative capabilities of GFlowNets, this framework proposes multiple candidate actions at each decision point, enabling robust personalization across heterogeneous user setups and network environments. The experimental results demonstrate that our approach outperforms rule-based and reinforcement learning-based methods in terms of Quality of Experience and bandwidth usage.

Beyond the specific application of short video feeds, we outline a unified GFlowNet-based framework generalizable to broader multimedia system contexts. Our work takes a significant step toward reimagining personalization through multi-candidate generative modeling and opens the door to integrating GFlowNets with real-world multimedia infrastructures, adaptive user modeling, and scalable, multi-objective optimization.

\bibliographystyle{ACM-Reference-Format}
\bibliography{ref}

\end{document}